# Life, the Universe, and almost Everything: Signs of Cosmic Design?

Rüdiger Vaas

Center for Philosophy and Foundations of Science,
University of Giessen, Germany

Ruediger.Vaas@t-online.de

______________________________________

*Draft – critical comments welcome!*
______________________________________

**Abstract:**

*Why did the big bang occur, why do the laws and constants of nature as well as the boundary conditions seem so fine-tuned for life, what is the role of intelligence and self-consciousness in the universe, and how can it escape cosmic doomsday? The hypothesis of Cosmological Artificial Selection (CAS) connects those questions and suggests a far-reaching answer: Our universe might be understood in terms of vast computer simulations and could even have been created and transcended by one. – This essay critically discusses some of the premises and implications of CAS and related problems both with the proposal itself and its possible physical realization: Is our universe really fine-tuned, does CAS deserve to be considered as a convincing explanation, and which other options are available to understand the physical laws, constants and boundary conditions? Is life incidental, and does CAS revalue it? And is intelligence and self-consciousness ultimately doomed, or might CAS rescue it?*

> "Many worlds might have been botched and bungled, throughout an eternity, 'ere this system was struck out. Much labour lost: Many fruitless trials made: And a slow, but continued improvement carried out during infinite ages in the art of world-making."
> David Hume (1779)

"We are called to be architects of the future, not its victims", Buckminster Fuller was convinced. And Eric Hoffer claimed: "The only way to predict the future is to have power to shape the future". But the far future of our universe as well as its beginning raise deep and difficult questions (Vaas 2010a): How will it evolve and perhaps end, and why did the big bang occur in the first place? Furthermore, the possibility of life and intelligence is closely connected to these questions: Its final fate, if unchallenged, appears deadly dark (Vaas 2006a & 2010b), and its origin and continuation depends on specific boundary conditions as well as the laws and constants of nature, which seems to be special or extremely improbable (Vaas 2004a).

Clément Vidal (2010) scrutinized all those issues and connected them in an ambitious and speculative way. According to his proposal, the presumably fine-tuned laws and constants of nature can be interpreted as a result of *cosmological artificial selection* (CAS) – as if they were created either physically or within an advanced computer simulation. This *artificial cosmogenesis* could even save life from cosmic extinction if there is an escape into other universes, perhaps designed for that purpose, before ours is ultimately doomed. So according to Vidal the CAS hypothesis provides (1) an understanding of our apparently fine-tuned universe by explaining or reconstructing it with advanced simulations and as a possible result of cosmic design, (2) a fundamental role of life and intelligence in our universe and beyond, perhaps even refuting the impression that it is epiphenomenal, incidental, futile or absurd, and (3) a chance for a really long-term or even eternal existence of life by offering a way out of our universe if it is ultimately doomed. Put differently, the strength of CAS might consist in a better understanding of (1) the origin, (2) the meaning, and (3) the potential far future of life. These three issues are connected, but nevertheless logically and physically independent from each other – thus if one is wrong or inadequate, the other ones might still stand.

The following comments shall critically reflect those issues, investigating some of their premises and implications, and discuss the related problems both with the CAS proposal itself and its possible physical realization. First it shall be reconsidered whether our universe is really fine-tuned and whether CAS deserves to be considered as a convincing explanation. Second it shall be asked whether life is incidental and whether CAS revalues it. And third it shall be discussed briefly whether self-conscious, intelligent life is ultimately doomed and whether CAS might rescue it.

## 1    Is our universe fine-tuned?

Life as we know it depends crucially on the laws and constants of nature as well as the boundary conditions (e.g. Leslie 1989, Vaas 2004a, Carr 2007). Nevertheless it is difficult to judge how fine-tuned it really is, both because it is unclear how modifications of many values together might compensate each other and whether laws, constants and initial conditions really could have been otherwise to begin with. It is also unclear how specific and improbable those values need to be in order information-processing structures – and, hence, intelligent observers – to develop. If we accept, for the sake of argument, that at least some values are fine-tuned, we must ask how this can be explained.

In principle, there are many options for answering this question (table 1). Fine-tuning might (1) just be an *illusion* if life could adapt at very different conditions or if modifications of many values of the constants would compensate each other; or (2) it might be a result of (incomprehensible, irreducible) *chance*, thus inexplicable (Vaas 1993); or (3) it might be *nonexistent* because nature could not have been otherwise, and with a *fundamental theory* we will be able to prove this; or (4) it might be a product of *selection*: either *observational selection* within a vast multiverse of (infinitely?) many different realizations of those values (*weak anthropic principle*), or a kind of *cosmological natural selection* making the measured values (compared to possible other ones) quite likely within a multiverse of many different values, or even an *teleological or intentional selection*. CAS is one example of the latter – but there are also other alternatives, for instance some versions of the strong anthropic principle and even theistic or deistic creation. (There are further and even more bizarre options, like solipsism, but they shall be neglected here.) Even worse, those alternatives are not mutually exclusive – for example it is logically possible that there is a multiverse, created according to a fundamental theory by a cosmic designer who is not self-sustaining, but ultimately contingent, i.e. an instance of chance.

To summarize, the reasoning goes like this:

>Premise (1): There is a world w with some specific properties p.
>Premise (2): X explains p or w.
>Premise (3): There is X.
>Conclusion: p or w are explained by X

Here X stands for irreducible chance, for a fundamental or at least deeper theory, for a multiverse, for observational selection (weak anthropic principle), for cosmological natural selection, for cosmological artificial selection (cosmic engineers or simulation), for a kind of teleological anthropic principle (i.e. an impersonal teleological force or an intentional transcendent designer, e.g. god), or for a combination of more than one of these (e.g. observational and cosmological natural selection require the multiverse). The task for physics and cosmology is, thus, to find out whether those three premises are true and what p and X are.

From both a scientific and philosophical perspective the fundamental theory approach and the multiverse scenario are most plausible and heuristically promising (Vaas 2004a & 2004b).

*Table 1. Digging deeper: Laws, constants and boundary conditions are the basic constituents of cosmology and physics from a formal point of view (besides spacetime, energy, matter, fields and forces or more fundamental entities like strings or spin-networks and their properties with regards to content). An ambitious goal and at least historically a successful heuristic attitude is reduction, derivation and unification to achieve more fundamental, far-reaching and simple descriptions and explanations. While uniqueness is much more economical and predictive, multiple realizations – presumably within a multiverse – have recently be proposed as an opposing (but not mutually exclusive) tendency. This table provides a summary of different approaches, possibilities and problems; it is neither complete nor systematically unrivaled.*

|  | fundamental laws (L) of nature | fundamental constants (C) | boundary conditions (BC) |
|---|---|---|---|
| **uniqueness? (and just one universe?)** | (1) irreducible and disconnected?<br>(2) or derived from or unified within or reducible to one fundamental theory ("Theory of Everything", TOE)?<br>• "logically isolated"?<br>• or even logically sufficient (self-consistent)? (Bootstrap principle)<br>• ultimately (logically?) deducible? (thus without empirical content if analytically true? natural science as pure mathematics?)<br>(3) or nonexistent because emerging via self-organization from an underlying chaos ("law without law" approach) | (1) irreducibly many?<br>(2) or only a few or just one (e.g. string length)?<br>• with a unique value? (just random or determined by L?)<br>• with (infinitely?) many possible different values? (according to a probability distribution determined by L?)<br>(3) or ultimately none?<br>• because C just as conversion factors (e.g. in a TOE)<br>• and/or better understandable as initial conditions (with many different realizations in a multiverse?) | (1) as initial conditions?<br>• irreducible?<br>• not accessible? (due to inflation, mixmaster universe, BKL chaos, "cosmic forgetfulness", decoherence etc.)<br>• nonexistent (in eternal universe models)?<br>• explanatory irrelevant? (because convergence due to an attractor or replaced by present BC?)<br>(2) or as present conditions?<br>• sufficient?<br>• or the only useful ones? (e.g. as loop quantum cosmology constraints or according to the top down approach in Euclidean quantum gravity)<br>• or as (additional?) final conditions? (e.g. constraints in some quantum cosmology models)<br>(3) or nonexistent because determined by L? (e.g. the no-boundary proposal) |
| **multiverse?** | (1) many realizations<br>• as BC if not reducible<br>• separated in principle or with a common origin (cause)?<br>• restricted by or derived from a TOE, or truly random/irreducible?<br>(2) or with every (logically? metaphysically?) possible realization? (mathematical democracy, ultimate principle of plentitude)<br>(3) TOE as a "multiverse generator"? | (1) many realizations<br>• truly random<br>• restricted by L or BC?<br>(2) or every possible combination of values? (equally or randomly distributed?)<br>(3) or with some absolute frequency according to a probability distribution (determined by L, e.g. string statistics and/or within a framework like cosmological natural selection?)<br>(4) or just observational bias/selection (weak anthropic principle) from a random set? | (1) many realizations<br>• restricted variance due to an attractor (determined by L)?<br>• or truly random<br>(2) or every possible combination? (equally or randomly distributed?)<br>(3) or with some absolute frequency according to a probability distribution (determined by L and/or within a framework like cosmological natural selection)?<br>(4) or just observational bias/selection (weak anthropic principle) from a random set? |
| **design?** | • transcendent realization? (nonphysical causation)<br>• random creation or cosmological artificial selection by cosmic engineers?<br>• universal simulation/emulation? (or just a subjective illusion?)<br>*however:* design of one universe or of a multiverse? and what have caused the designer? are there even transcendent (Platonic) laws? why is there someone rather than no one? | | |
| **randomness?** | ultimately yes (even if there is only one self-consistent TOE)<br>• Gödel-Turing-Chaitin theorems | not necessarily (if determined by L or reduced to BC) | not necessarily (if determined by L) |

## 1.1 Theoretical explanation instead of fine-tuning

It is a very controversial issue whether and how far reductionism works in physics – and beyond. Higher-order levels of descriptions are undoubtedly necessary for practical purposes but might still be (approximately) reducible to lower levels (depending on certain constraints of course, i.e. boundary conditions). Putting such issues and the ambiguous meaning of "reduction" (Vaas 1995a) aside, one might argue roughly like this: The bedrock of reality consists of matter-energy, interactions, and spacetime (or even more fundamental "building blocks" like loops or strings), the properties, states and dynamics of which can be described by what it is called *laws and constants of nature* and a set of *initial or boundary conditions*. For explanations, these descriptions, embedded and joined in a theory, should suffice in principle. Thus, the usual scheme of explanations in physics is roughly like this:

Given some boundary conditions and laws (including constants) or a theory (which connects or unifies different laws), i.e. the e*xplanans*, some facts or events, i.e. the *explanandum*, can be explained (*retrodiction*) or forecasted (*prediction*) and thereby tested. Different kinds of scientific explanation (e.g. Hempel 1965, Pitt 1988, Salmon 1998, Woodward 2003/2009, Lipton 2004, Mayes 2005) are within this scheme, e.g. the deductive-nomological explanation (covering law model) with deterministic laws, the inductive-statistical explanation with probabilistic laws (but with any probability?), and the causal explanation focusing on cause-effect-relations (which might be either deterministic or probabilistic).

This scheme works pretty well. However, one can still ask: Why those laws (or theory, respectively), why those constants, and why those boundary conditions? If our universe is not eternal, or at least not its laws and constants, these questions are related to another, namely: What is the explanation for the big bang?

But these questions are different compared to the usual ones regarding physical explanations, because they already presuppose or contain what should be explained. Take the big bang for instance, i.e. the hot and dense very early universe. It is described by observations (e.g. the expansion of space, the cosmic background radiation and its properties, the ratios of light elements etc.) and laws or theories (especially general relativity, thermodynamics, high-energy physics). But neither the big bang, nor those laws and theories, are explained (retrodicted) by those observations. The observations are explained by big bang theory, not vice versa. So how to explain the big bang, i.e. how did the hot and dense state of the very early universe come into existence? Here, new theories (or constraints of the current theories) and data are required. Some even argue that a new scheme of explanation is needed – perhaps an anthropic, functional, teleological, or even transcendent one? This would be one of the largest paradigm changes since the advent of classical physics. Insofar as CAS constitutes a certain flavor of such a new kind of explanation, neither its challenge nor the reluctance against it should be underestimated.

It is therefore wise to push the ordinary explanation scheme of physics as far as possible and see how far one might really get. Thus the explanandum now is the big bang and its (causal) connections to the present/observable universe. And the question is which explanans might suffice: which fundamental laws (e.g. of M-theory, supersymmetric grand unified theories, general relativity, etc.), fundamental physical constants (e.g. c, h, G), and initial conditions (e.g. dimensionality, metric, values of the fundamental fields, fluctuations etc.)?

Furthermore, one can ask whether there is a way to simplify the "triangle" of laws, constants and boundary conditions, i.e. to reduce one of its "corner" – or even two – to another one (Table 1). This would be a huge breakthrough in physics. Different possibilities are under consideration, but this is more or less pure speculation yet:

• Constants might be reduced to boundary conditions: For instance a huge "landscape" of solutions in string theory could exist, depending on different compactifications of tiny extra dimensions etc. (Douglas 2010); if all these mathematical solutions are (or could be) physically realized, e.g. as bubble universes in the exponential growing false vacuum of eternal inflation, then those boundary conditions, set by the phase transition to a specific "true" vacuum of an originating bubble universe, might determine what appears as constants of nature in such a universe (Linde 2005 & 2008, Aguirre 2010; for an estimate of the gigantic number of different universes in the multiverse see Linde & Vanchurin 2009). Another and not mutually exclusive scenario is cosmological natural selection (see below).

• Laws might also be reduced to boundary conditions: Cosmological natural selection is an example here again, at least for some laws. And if no law and constant describing our universe is fundamental, but all are ultimately random, they can also be seen as boundary conditions in a wider sense. Thus, a specific set of laws might just be a set of boundary conditions with respect to a specific (kind of) universe within a multiverse. However, this does not necessarily imply a "law without law" approach (Wheeler 1980 & 1983) in the strict sense – i.e. that the only law is that there are no (fundamental) laws, but pure randomness – because there might be a fundamental (although contingent) law nevertheless, which rules the multiverse-generating processes.

• Boundary conditions might, on the other hand, be reduced to laws: A famous example is the no-boundary proposal in quantum cosmology (Hartle & Hawking 1983, Vaas 2008a).

Up to now this discussion was quite abstract, surveying a range of possibilities. But there is something more to say which might add some flesh (albeit still not very nutritious yet) to the backbone of particle physics and cosmology. Their standard models contain at least 31 free parameters which have to be measured and cannot be explained yet (Tegmark et al. 2006, Tegmark 2010). Perhaps more advanced models or theories will reduce the number of those parameters significantly. But there is no guarantee for this. More fundamental approaches like supersymmetry might even increase that number, and perhaps further theories are needed with their own constants. However, string theory deigns to provide only very few, perhaps only two (Duff et al. 2002): the string length and the speed of light (*pacem* critics who claim that string theory rather dangles on a string). Sometimes it was argued that the number of dimensional constants like the gravitational constant G, the speed of light c or the reduced Planck constant $\hbar = h/2\pi$ (with dimensions $m^3 \cdot kg^{-1} \cdot s^{-2}$, $m \cdot s^{-1}$ and $m^2 \cdot kg \cdot s^{-1}$ respectively) could be dropped (normalized to 1). They depend on conventions, i.e. our unit system, and are dimensional again in other unit systems – and such dimensions are necessary for making measurements. Dimensionless constants, however, are pure numbers and independent of any unit system. They are ratios between physical quantities, e.g. the electron-proton mass ratio $m_e/m_{pr} \approx 1/1836.15$.

Often in calculations the units of fundamental constants are normalized to 1 (e.g. $G = c = \hbar = 1$). But this is only a simplification. To make measurements, the units are still relevant. However, one can go a step further and define "natural" units independent from human standards, i.e. as dimensionless constants. Indeed, h, c and G can be seen as mere conversion factors. c transforms energy into mass ($E = mc^2$), h energy into frequency ($E = h\nu$) and G mass into length, namely into the Schwarzschild radius $R_s$, the radius of a black hole ($R_s = 2GM/c^2$). Taking another important quantum property into account, the Compton wavelength ($\lambda_c = h/mc$), the following is possible: a mini black hole with the Compton wavelength as its Schwarzschild radius can act as a natural measuring rod and thermometer as well as clock and weighing machine (Duff et al. 2002). Any extraterrestrial civilization could understand it. But doing without any arbitrary human scales and definitions, "true" constants must be expressed with pure numbers, i.e. independent of dimensional quantities like velocity, mass and length or reference systems like black holes. This is not possible with Planck units alone. To get pure numbers one has to multiply them with another dimensional constant, e.g. the proton mass $m_{pr}$. For example one gets $m_{pr}^2 G/\hbar c = 10^{-38}$. Every physicist in our universe could understand, measure and use such a value, independent from its yard stick, and discuss it with any other habitant in any other galaxy. But of course the question remains why this value is $10^{-38}$ and not something else.

Future theories of physics might reveal the relations between fundamental constants similarly to what James Clerk Maxwell has shown by unifying electric and magnetic forces: that three until then independent constants – the velocity of light c, the electric constant $e_o$ (vacuum permittivity), and the magnetic constant $\mu_o$ (vacuum permeability) – are connected with each other: $c = (\mu_o \cdot e_o)^{-0,5}$. Indeed some candidates for a grand unified theory of the strong, weak and electromagnetic interaction suggest that most of the parameters in the standard model of particle physics are mathematically fixed, except for three: a coupling constant (the electromagnetic fine-structure constant) and two particle masses (namely that of down and up quarks) (Hogan 2000). A promise of string theory is even to get rid of any free parameter – if so, all constants could be calculated from first principles (Kane et al. 2002). However, this is still mainly wishful thinking at the moment. But it is a direction very worth to follow and, from a theoretical and historical point of view, perhaps the most promising.

So even without an ultimate explanation fine-tuning might be explained away within a (more) fundamental theory. Most of the values of the physical constants should be derived from it, for example. This would turn the amazement about the anthropic coincidences into insight – like the surprise of a student about the relationship $e^{i\pi} = -1$ between the numbers e, i and $\pi$ in mathematics is replaced by understanding once he comprehends the proof. Perhaps the fact that the mass of the proton is 1836 times the mass of the electron could be similarly explained. If so, this number would be part of the rigid formal structure of a physical law which cannot be modified without destroying the theory behind it. An example for such a number is the ratio of any circle's circumference to its diameter. It is the same for all circles in Euclidean space: the circular constant $\pi$.

But even if all dimensionless constants of nature could be reduced to only one, a pure number in a theory of everything, its value would still be arbitrary, i.e. unexplained. No doubt, such a universal reduction would be an enormous success. However, the basic questions would remain: Why this constant, why this value? If infinitely many values were possible, then even the multitude of possibilities would stay unrestricted. So, again, why should such a universal constant have the value of, say, 42 and not any other?

If there would be just one constant (or even many of them) whose value can be *derived* from first principles, i.e. from the ultimate theory or a law within this theory, then it would be completely explained or reduced at last. Then there would be no mystery of fine-tuning anymore, because there never was a fine-tuning of the constants in the first place. And then an appreciable amount of contingency would be expelled.

But what would such a spectacular success really mean? First, it could simply shift the problem, i.e. transfer the unexplained contingency either to the laws themselves or to the boundary conditions or both. This would not be a Pyrrhic victory, but not a big deal either. Second, one might interpret it as an analytic solution. Then the values of the constants would represent no empirical information; they would be no property of the physical world, but simply a mathematical result, a property of the structure of the theory. This, however, still could and should have empirical content, although not encoded in the constants. Otherwise fundamental physics as an empirical science would come to an end. But an exclusively mathematical universe, or at least an entirely complete formal description of everything there is, derivable from and contained within an all-embracing logical system without any free parameter or contingent part, might seem either incredible (and runs into severe logical problems due to Kurt Gödel's incompleteness theorems) or as the ultimate promise for the widest and deepest conceivable explanation. Empirical research, then, would only be a temporary expedient like Ludwig Wittgenstein's (1922, 6.54) famous ladder: The physicist, after he has used empirical data as elucidatory steps, would proceed beyond them. "He must so to speak throw away the ladder, after he has climbed up on it."

## 1.2 Natural selection instead of fine-tuning

What appears fine-tuned might be not – either because it is a unique, derivable consequence of an underlying lawful structure, and hence determined, or because it is the probable outcome of a stochastic process. An especially attractive possibility, also from an explanatory perspective, is a kind of Darwinian evolution of the values of fundamental constants (and perhaps even laws and boundary conditions). As in biology, i.e. evolutionary theory, ostensible features of design would be revealed as results of a nonintentional, self-organized process based on mutation, selection and differential reproduction. Darwinian evolution is a well-established, indeed paradigmatic case of such a "blind" self-organization leading to astonishingly complex structures (Dennett 1995, Kanitscheider 2009, Vaas 2009a). It is therefore a reasonable, although bold speculation to blow up a Darwinian kind of explanation to a cosmological scale within a multiverse framework.

In contrast to observational selection or bias according to the weak anthropic principle, which works in any multiverse scenario, but is not predictive, an observer-independent selection mechanism must generate unequal reproduction rates of universes, a peaked probability distribution or another kind of differential frequency. For example, as Andrei Linde (1987) has pointed out firstly, the constants of nature might vary from one inflationary domain to another, generating different rates of exponential expansion and bubble universe formation.

Up to now the most elaborated model of cosmological natural selection is Lee Smolin's scenario (1992, 1997 & 2010). Actually it has coined the whole approach, including its name. In contrast to observational selection, Smolin's scenario of cosmological natural selection (CNS) is predictive and, thus, directly testable and falsifiable – at least within certain assumptions. (Note that CNS implies observational selection, but not vice versa.)

The hypothesis of CNS assumes, like CAS and anthropic observational selection, the existence of a multiverse or of a landscape of possible low energy parameters. Furthermore, CNS assumes that black hole interiors bounce and evolve into new universes; that the values of the fundamental parameters can change thereby in small and random ways; that, therefore, different universes have different reproduction rates – universes with more black holes create more offspring universes; and, hence, that it is very probable after sufficient time that a universe chosen at random from a given collection of physically possible universes has parameters that are near a maximum of black hole production. If our universe is a typical member of that collection, then its fundamental parameters must be close to one of the maxima of the black holes production rates. Hence, our universe is selected for maximizing its number of black holes, and it is a descendant of universes which were already selected for this. Therefore, the fundamental parameters have the values we observe because this set of parameters leads to a (local) maximum of descendant universes, making the production of black holes much more likely than most alternatives. Thus, there is no need for invoking the weak (or even strong) anthropic principle – the existence of life is not used as part of the explanation of the parameter values. Contrariwise, its preconditions can be explained by CNS because the existence of stars and, as an offshoot, carbon chemistry, comes along with an efficient black hole formation rate. Therefore it is possible to continue physical research within a multiverse scenario without invoking the anthropic principle. In particular, this is true whether or not the ensemble of universes generated by bouncing black holes is a sub-ensemble of a larger ensemble that might be generated by a random process such as eternal inflation. And CNS leads to a testable prediction: Most changes in the fundamental parameters would decrease the rate at which black holes are produced in our universe or leave it unchanged, but would not increase it. This prediction still holds.

So CNS is quite successful from a theoretical point of view. Because this weakens the prospects for CAS, some crucial open issues and problems for CNS should be analyzed more closely now (see Vaas 1998 & 2003). If CAS could do better here, this would be a huge achievement.

First, there are open questions regarding new universes emerging out of black holes. Even if black holes are places of birth for universes it is not clear whether the values of the physical parameters really vary by *small* amounts and *randomly* as it is presumed, what happens to the already born universes if their mother black holes merge together or evaporate, whether there are further universes created if black holes merge together (and how many: one or two?), why there is only *one* offspring from a black hole and not (infinitely) many, and, if the latter is true, whether the numbers of new universes which are born from each black hole may differ according to the mass of the black hole. It was suggested that a large number of universes might be created inside each black hole and that the number of universes produced that way may grow as the mass of the black hole increases (Barrabès & Frolov 1996). If so, universes should be selected for *supermassive* black holes, not for sheer numbers of black holes.) It is also not clear whether the different universes interact which each other. There is even the threat of a reciprocal destruction. Another restriction of Smolin's approach is that his cosmic reprocessing mechanism only leads to different values of parameters, but not to different *laws*. His hypothesis still requires the same basic structure of the laws in all the universes. But of course an even more radical proposal – a variation not only of constants but also of laws – is beyond the possibility of scientific investigation (at least for now). We simply do not know whether a distinction is useful between universes which are physically possible, as opposed to those that can only be imagined (which are only metaphysically or logically possible).

Second, there is no necessary connection between black holes and life. In principle, life and CSN could be independent of each other. There are two reasons for this: On the one hand there may be universes full of black holes were life as we know it couldn't evolve. For instance it might be possible that there are only short-lived giant stars which collapse quickly into black holes, or that there are universes dominated either by helium or by neutrons (corresponding to the neutron/proton mass difference being either zero or negative), or that there are universes without stars at all but many primordial black holes. Such universes might be very reproductive because of their giant stars or primordial black holes but are not able to produce earth-like life. On the other hand we can conceive a universe without black holes at all (if supernovae lead to neutron stars only) but which could be rich in earth-like life nevertheless. Thus, there is not a (logically) necessary connection between black holes and life (Rothman & Ellis 1993, Ellis 1997). Smolin's CNS hypothesis therefore cannot necessarily explain the presumed fine-tuning of the universe for life. By the way, Smolin (1997, p. 393) also stressed that both properties of our universe – containing life and producing a maximum number of black holes – must be taken as independent for the purpose of testing the theory.

Nevertheless there may be a contingent connection between black holes and life – via the role of carbon as the "molecule of life", because its ability to make complex molecules (to a much larger extent than any other element), and as an element accelerating star formation, because of the role that carbon monoxide plays in shielding and cooling the giant molecular clouds of gas and dust were stars are born (Smolin 1997). Thus, there may be at least a coincidence between the conditions for maximizing black hole formation and being hospitable for life. But we must still wonder why the laws of nature are such that this linkage occurs. (And it is not clear, whether carbon monoxide really does increase the number of black holes, because hydrogen cools efficiently, too, and all the stars in the early universe were carbon-free.)

Third, in CNS life and intelligence are a kind of epiphenomenon – contrary to CAS. If they do not contribute to black hole formation, life and intelligence are a mere by-product in Smolin's scenario, i.e. they are causally inert (regarding the evolution of the universes). Thus, in CNS our universe was not positively selected for life, even if the conditions of life would be exactly the same as the conditions for maximizing black holes. Therefore, CNS does not imply (or entail) a "meaning", function or advantage of life. But couldn't it be possible, that, nonetheless, there is a hidden connection between the hospitality of universes for life on the one hand and black hole formation on the other? Perhaps black holes could be advantageous for life, or life could be advantageous for black hole formation. Therefore a CAS proponent can argue that cosmic engineers might create universes by means of black holes (see below). There would be no self-organized evolution in this case but rather a preplanned development. Nevertheless, life could still be seen as a "tool" of the multiverse to produce more universes. But it would be no epiphenomena in this case.

Fourth, there are problems with predictability and testability. According to Smolin, there are eight known variations in the values of fundamental parameters that lead to worlds with fewer black holes than our own, but there is no variation known with has clearly the opposite effect. This is already an interesting observation. Furthermore, Smolin's hypothesis has some predictive power, because there are physical effects and properties influencing black hole formation rates which are still not known (at least not precisely enough). Here, Smolin's hypothesis provides some constraints for these effects if the number of black holes is almost maximized. These predictions can be tested in principle and partly even in the near future. They are related to the masses of $K^-$ mesons (and hence the mass of the strange quark), electrons and neutron stars, the strength of the weak interaction, the density of protons and neutrons, the duration of the presumed inflationary epoch of the early universe, temperature patterns of the cosmic microwave background, the black hole formation rate dependent of the gravitational constant, etc. These predictions are testable in principle, and they are at home in the reign of current cosmology and particle physics. However, Smolin's central claim cannot be falsified. Falsifiability of a hypothesis depends on holding fixed the auxiliary assumptions needed to produce the targeted conclusion. In practice, one tries to show that the auxiliaries are themselves well confirmed or otherwise scientifically entrenched. What should be falsifiable according to Smolin is his claim that *our* universe is nearly optimal for black hole formation. However, this is not a necessary consequence of his premises. A consequence is only that *most* universes are nearly optimal. To move from this statistical conclusion to the targeted conclusion about our universe, Smolin (1997, p. 127) simply assumes that our universe is typical. This is an additional hypothesis as he admits. But this auxiliary assumption is neither confirmed nor otherwise scientifically entrenched. Thus, if changes in the values of our parameters did not lead to a lower rate of black hole formation – contrary to Smolin's prediction – we could always "save" CNS by supposing that our universe is not typical. Hence, there is (at least at the moment) no possibility to falsify Smolin's central claim that our universe is nearly optimal for black hole formation (Weinstein & Fine 1998). One could introduce other ad hoc hypotheses as well in order to keep the central idea of CNS. But this might undermine its falsifiability. For example, if there are parameters whose variation from their actual value leads to an increase of black hole formation, one could still claim that these parameters cannot be varied without also changing other parameters, leading e. g. to large side-effects in star formation, hence the originally varied parameters are no *independent* parameters contrary to the assumption. But note: if there is a Theory of Everything someday, which uniquely determines the values of the physical parameters, the CNS hypothesis would still be falsified at last.

Fifth, Smolin's Darwinian analogy of CNS is in some respect misleading. Natural selection as described in biology depends on the assumption that the spread of populations (or genes) is mainly constrained by *external* factors (shortage of food, living space, mating opportunities etc.). In comparison with that, the fitness of Smolin's universes is constrained by only one factor – the numbers of black holes –, and this is an *internal* limitation. Furthermore, although Smolin's universes have different reproduction rates, they are not competing against each other (Maynard Smith & Szathmáry 1996). Although there are "successful" (productive) and "less successful" universes, there is no "overpopulation" and no selective pressure or "struggle for life", hence no natural selection in a strict biological (Darwinian) sense. Smolin's universes are isolated from each other (except maybe for their umbilical cords). Therefore, there couldn't be a quasi-biological evolution of universes.

Thus, a central feature of Smolin's CNS scenario is that the values of the constants were not selected due to competition but only due to differential reproduction: Some universes have more offspring than others, but there is no rivalry about resources, space etc. as in life's evolution according to Darwinian natural selection. However one could envisage other scenarios of cosmic evolution where not only mutations of natural constants occur, leading to differential reproduction, but also competition between the universes or their origins and, hence, a Darwinian selection process. (Such descriptions are analogous to those of biological evolution, but of course do not refer to that in the strict sense; for the heuristic value and advantages of analogies see Vidal 2010.)

For instance, universes might nucleate out of accidental fluctuations within a string vacuum (Gasperini & Veneziano 2010) or within the vast landscape of string theory (Douglas 2010), inheriting certain properties. There could be a natural selection of such universes depending, for example, on their energy and matter densities (cf. Mersini-Houghton 2010): If the matter density is above a certain limit (or the cosmological constant is negative), the emerging universe would rapidly collapse and vanish; other universes would expand too fast, if their vacuum energy (cosmological constant) is too large – matter or structures like stars and, thus, life could not form in it. If the emergence of such universes would affect (suppress? increase?) the formation probabilities of others, for example by influencing their "surrounding" parts of the string vacuum or landscape, a kind of competition could be the result.

Another example of cosmic Darwinism was recognized in the stochastic approach to quantum cosmology (García-Bellido 1995): In Brans–Dicke chaotic inflation, the quantum fluctuations of Planck mass $M_p$ behave as mutations, such that new inflationary domains may contain values of $M_p$ that differ slightly from their parent's. The selection mechanism establishes that the value of $M_p$ should be such as to increase the proper volume of the inflationary domain, which will then generate more offspring. (Of course this selection mechanism only works if the values of the fundamental constants are compatible with inflation.) It is assumed here that the low energy effective theory from strings has the form of a scalar-tensor theory, with nontrivial couplings of the dilaton $\phi$ to matter. It is therefore expected that the description of stochastic inflation close to Planck scale should also include this extra scalar field. Brans–Dicke theory of gravity is the simplest scalar-tensor theory, containing a constant $\omega$ parameter. The string dilaton plays the role of the Brans–Dicke scalar field, which acts like a dynamical gravitational coupling: $M_p^2(\phi) = 2\pi\phi^2/\omega$. This scenario is in principle testable, by the way, because it predicts that the larger $M_p$ is in a given inflationary domain, the smaller the amplitude of density perturbations should be. The universe evolves towards largest $M_p$ and smallest amplitude of density perturbations compatible with inflation, which agrees well with observations. Thus our universe, with its set of values for the fundamental constants, would be the offspring of one of such inflationary domains that started close to Planck scale and later evolved towards the radiation and matter dominated eras.

It is not that important here whether one of the sketched scenarios turns out to be true. The point here is their general idea. What it shows already is that cosmological natural selection provides a quite simple physical explanation of what seems as mysterious fine-tuning – an explanation without any reference to intentionality or design. Analogous to Darwinian evolutionary theory in biology, the apparently sophisticated structure of the foundations of our universe might simply be the result of a multiversal self-organization. This is a parsimonious and straightforward explanation. Whether it is true is not a philosophical question however, but depends on empirical and theoretical data – as the other main approach discussed above: the hypothesis that the fine-tunings can be derived from a fundamental law. Nevertheless, a philosophical advantage should not be neglected. So why work out another scenario, cosmological artificial selection? Can it do even better?

## 1.3 Artificial selection as fine-tuning

### 1.3.1 Creation out of something

A model of cosmological artificial selection aims to explain the presumed fine-tuning of our universe easily: As the result of a goal-oriented, intentional action. This might or might not lead directly to our universe, and it might entail a certain amount of randomness (see below). But artificial cosmogenesis (Vidal 2008 & 2010) seems to be at least a possibility to enhance or alter the natural selection of real universes. This might be done via studying and selecting simulated universes first, investigating the range of physical possibilities, or it might be done directly via making or starting new universes (see below). Note that artificial selection in biology, on animals or plants or microorganisms, does not replace natural selection, and it does not "design" new organisms or create them from scratch; it tries only to foster some traits over others. So CAS might just extend this manipulation to cosmological scales. But if the new universes are meant as new homes for their creators, because the universe they live in will run out of free energy and life-friendly conditions, the laws and constants of those successor homes will probably be intended to remain fixed – otherwise the cosmic engineers would die after moving in. So an important part of CAS might consist in carefully selecting the right conditions, perhaps via extensive computer simulations prior to the replication events, and therefore the successor universes would really be the result of an intentional fine-tuning.

Note that CAS might be realized both without a fundamental theory or by means of it. If there is a unique set of laws and constants with no alternatives, it might allow the creation of new universes nevertheless and, possibly, some variations of initial conditions. However, is CAS more convincing than multiverse scenarios doing without intentional causation? Or, to put it differently, why should a multiverse scenario not suffice?

Neglecting the possible meaning and far future of life for the moment, CAS has to be defended against two much simpler kinds of multiverse scenarios: (1) those with a random distribution of laws, boundary conditions and values of constants, and (2) those with a probability distribution of some kind.

Within the first scenario, the fine-tunings can be understood by anthropic bias or observational selection, i.e. the weak anthropic principle: We are able to live only in a specific universe whose physical properties are consistent with our existence – a prerequisite for it –, and therefore we should not be surprised that we observe them. Thus according to the weak anthropic principle, the world consists of an ensemble of universes, a multiverse, with different laws and parameters of nature, and we can detect only those which are compatible with our existence or even among its necessary conditions. But this is, strictly speaking, not a physical explanation (Vaas 2004a) and might not even be testable, i.e. predictive (Smolin 2010). So proponents of CAS could claim that CAS is superior.

Within the second scenario, the fine-tunings can be understood either via a case of cosmological natural selection or as the result of an underlying law, determining a probability distribution with a (local or global?) maximum, and the physical properties of our universe are in the vicinity of this maximum, i.e. quite probable. At the moment we do not know of such law. We might just assume there is one. A stronger position is to adopt the principle of mediocrity (Vilenkin 1995), saying that we are typical observers in a certain sense. However, this principle might not be applicable here; or our universe is special, i.e. not at or near the maximum of the probability distribution. Then we have to find a different explanation, or we must come back to anthropic observational selection.

Another option – compatible with more fundamental laws, perhaps even depending on it – is cosmological natural selection. It could be seen as both the nearest relative of and strongest alternative to CAS. It seems to be testable, but CAS proponents might still argue that it does not explain enough and has many problems. This is true at least for Lee Smolin's version, because the bounce within black holes and the physical "mutations" are still completely mysterious, and a "selection" mechanism is also missing.

### 1.3.2 Deism in the lab

From a speculative point of view, CAS might be praised for stressing that in principle design is – although not mandatory of course – at least possible within a cosmological and naturalistic framework. In contrast to theistic postulates of transcendent, nonphysical entities and causation, a CAS scenario is fully reconcilable with *ontological naturalism*. Cosmic engineers are not something "above" or "beyond" nature, i.e. the multiverse, but a part and, indeed, a product of it.

Of course the term "cosmic engineers" (Harrison 1995) is somewhat metaphorical. It indicates correctly that there must be an advanced technological activity at work. But what kind of technician or civilization this is supposed to be remains completely open. Perhaps the creators are organic or robotic individuals, perhaps it is a collective intelligence with a single (even nonpersonal?) mind, perhaps it pervades its universe completely or hides within castles made from neutron stars – most probably it radically exceeds our imagination. In some respects those "cosmic engineers" might be seen as god-like. But nevertheless they are not supernatural, not independent of spacetime and energy, not beyond the physical realm. They are "transcending" our universe, but not the multiverse. Thus, CAS is compatible with and part of ontological naturalism. It does not require new metaphysical entities or forces. And it does not contradict the scientific attitude – in fact it pushes it to the extreme. Therefore, in the CAS scenario "creation" does *not* mean theistic creation! CAS can be seen as a kind of creation out of something – in contrast to a divine creation out of nothing, a world-making *ex nihilo*, a Kabbalahistic tzimtzum, a mystical emanation or a mythical transformation of chaos into order. And artificial cosmogenesis can, in principle, be understood in physical or naturalistic terms entirely; no religious context is attached here.

From a theological perspective, CAS might be seen as a technocratic successor of creation myths nevertheless, a naïve secular belief, an exuberant scientism gone wild. This is not amazing. However, it goes astray. CAS is a scientific and philosophical hypothesis or speculation, not a substitute religion. CAS might be bold or beyond belief, depending on personal taste and sincerity, but it does neither attack the nature of science nor the science of nature. (Of course CAS proponents have to carry the burden of proof and should provide theoretical and empirical evidence, not the sceptics.) One could even think about some sort of naturalizing the divine – CAS as *deism in the lab*: If one defines deism simply as belief in a transcendent entity, absent any doctrinal governance, who created our universe but does not interfere with it anymore via miracles etc., the cosmic engineer(s) could be identified with such a deity, a "supreme being", "divine watchmaker", "grand architect of the universe" or "nature's god". Of course, this "god" is *not* the theistic one, it "transcends" *not* nature in general but only our universe, and "creation" is *not* a nonphysical causation. However, as classical deism claims, such an engineer god might indeed "be determined using reason and observation of the natural world alone, without a need for either faith or organized religion" (Wikipedia). By the way, even deistic interventions in human affairs (or with respect to our universe as a whole after its fabrication) are not excluded in principle if the grand architect is able to mesh with it for instance via gravitons from a fifth-dimensional bulk space or nonlocal quantum entanglements or wormholes (which simple-minded beings like humans might understandably perceive or interpret as miracle or revelation). And of course a CAS-like deism would be a truncated deism, because the religious versions of deism – and there is plenty of variation here – have a very different background and goal, and they often contain much more, including moral and spiritual aspects (e.g. Waring 1967, Gay 1968, Byrne 1989, Johnson 2009).

Admittedly, all this sounds much more like science fiction or science fantasy than serious science, and it was not put forward by CAS proponents. But if one wants to adopt a theological perspective here at all, from that point of view CAS has indeed something provocative to offer: a radically physical deistic natural theology. Said with a twinkle in the eye, CAS both puts deism in the lab (of physical and philosophical reasoning) and is a result of deism in the lab (as a presumable process somewhere in the multiverse).

Sure enough, such theological contexts or connotations reinforce doubts about CAS and show how what delicate issues this proposal raises. Critics might object that CAS is creationism or intelligent design in new clothes (and in certain respects it actually is); or that CAS reintroduces the teleological thinking that was painstakingly expelled in the history of physics and biology (and in certain respects it actually does); and that CAS blurs the distinction between science and theology/religion/metaphysics (which might also be the case if supernaturalism is watered down or abandoned). Such criticism might be exaggerated, but should be taken seriously. Therefore CAS proponents must emphasize the hypothetical character of their proposal as well as their own scientific (and even naturalistic?) stance; they must search for demarcation criteria between science and theology or metaphysics and accept them; they have to seek for rigorous testing, both theoretical and empirical; they should clearly stress the distinction between CAS and ideological creationism; and they should point out that cosmic engineers are not divine beings to worship or to suppliantly submit to. CAS is far from proven true and poses many crucial questions and problems, but as a cute hypothesis it deserves unprejudiced discussion like any serious effort to improve the understanding of the strange world we live in.

### 1.3.3   Objections and challenges

Sure enough, CAS has problems on its own.
First and foremost, there is the difficulty of realizing CAS: It is completely unclear whether universes can not only be simulated to some extent but also physically instantiated. A few scientific speculations are already on stage (see below), but still in their infancy.
Second, one must distinguish between intentional creation and simulation (even if it would be empirically impossible to decide between them from "within"). A simulated universe does not have all the properties of a physically real universe – as a simulated river might obey hydrodynamical equations but doesn't make anything wet. Admittedly, deep epistemological problems are lurking here. And perhaps it will be possible to make the simulation so real that the missing properties are simply irrelevant; or to make it at least so useful that, for instance, conscious life within it is possible and the creators could "upload" their minds, knowledge and experiences, surviving within their simulation if they can no longer do in their own universe. But the hardware problem remains: How can something simulate something else which is comparably complex? And if the programmer's universe is doomed, their universal computer and, hence, computed simulation sooner or later should be doomed too. – So perhaps we live in (and are!) a computer simulation (Bostrom 2003). But this might have implications that could even lead to a reductio ad absurdum. As Paul Davies (2007, p. 496) emphasized, "there is no end to the hierarchy of levels in which worlds and designers can be embedded. If the Church-Turing thesis is accepted, then simulated systems are every bit as good as the original real universe at simulating their own conscious sub-systems, sub-sub-systems, and so on *ad* *infinitum*: gods and worlds, creators and creatures, in an infinite regress, embedded within each other. We confront something more bewildering than an infinite tower of virtual turtles: a turtle fractal of virtual observers, gods and universes in limitlessly complex inter-relationships. If *this* is the ultimate reality, there would seem to be little point in pursuing scientific inquiry at all into such matters. Indeed, to take such a view is as pointless as solipsism. My point is that to follow the multiverse theory to its logical extreme means effectively abandoning the notion of a rationally-ordered real world altogether, in favour of an infinitely complex charade, where the very notion of 'explanation' is meaningless."

Third, there is a crucial questions: If there are cosmic engineers at work, perhaps some of them having fine-tuned our universe, how did they emerge in the first place? In other words: *What or who created the creator(s)?* – To avoid an infinite explanatory regress, it seems most probable that they arose naturally in a life-friendly universe themselves. But this shifts the problem, because at least the creator's home universe should have formed without any intentional fine-tuning. Thus, either its origin was pure chance or the outcome of cosmological natural selection or the result of a multiverse "generator" according to some fundamental laws etc. Therefore we are back at the beginning, i.e. the original question regarding fine-tuning. If our universe was created according to CAS, the fine-tuning problem is just shifted to the problem of explaining an earlier fine-tuned universe where the cosmic engineers evolved. Their home universe might have been physically simpler than ours, but not too simple either, otherwise such complex creators could never have emerged. So this is a major objection against CAS.

And, connected with it, there is a further problem: One might wonder whether CAS has any convincing explanatory force at all. Because ultimately CAS tries to explain something complex (our universe) with something even more complex (cosmic engineers and engineering). But the usual explanatory scheme is just the converse: The *explanans* should be simpler than the *explanandum*. Furthermore, CAS adds something qualitatively new: While multiverse (including CNS) and fundamental law approaches to the fine-tuning problem postulate some new nomological regularities, CAS postulates an intentional cause *in addition*. CAS is therefore a mixture of explanations: physical and intentional. (Intentions are not, as such, nonphysical, and actions can be conceptualized as causes – as specific causes, to be more precise (Davidson 2001) –, so *pacem* other opinions there is no reason to abandon naturalism here, but intentional explanations are nevertheless not epistemologically reducible to physical explanations.)

These arguments are not a knock-out objection. But they point out some severe difficulties with CAS. At least they show that CAS cannot be the ultimate explanation – like any other design scenario (Vaas 2006b). However, this is not what CAS proponents (should) have in mind anyway. And if it would be possible for us to carry out artificial cosmogenesis by ourselves, a strong case for CAS can be made even within its explanatory restrictions. And from a philosophical and practical perspective, CAS might be very important indeed.

## 2  Is life incidental?

One of the most remarkable developments in human cultural history was the recognition of our tiny place in the vast universe (or perhaps multiverse), and that we are not obviously meant to be here. The overcoming of a naïve and infantile anthropocentrism, that the universe is there for us, and, strangely connected, that an all-compassing god is there for us too (and vice versa) was one of man's great – and still not fully accomplished – achievements: an "emergence from his self-incurred immaturity" (Immanuel Kant). Darwinian theory of evolution suggested that man, and indeed life itself, was not ingeniously designed, but a result of self-organizing processes, a "fruit of chance and necessity", as Jacques Monod (1970) used to cite Democritus. Astrophysics, big bang theory and, finally, the still speculative scenarios of quantum cosmology taught the same lesson albeit on much larger scales: The emergence of intelligence was more or less an accident, not meant to be there in a universe that is indifferent to life's concerns, goals and values. However, in intelligent, self-conscious beings like humans the universe at least became partly aware of itself, poetically speaking.

But self- and I-consciousness also revealed the absurdity of life in the face of chance, futility and misery (Vaas 1995b, 2008b & 2009b). The shirking of belief in transcendent creators or in an almighty, omnibenevolent god, though perhaps consolatory for some (Vaas 2009c), cannot surmount absurdity because misery, injustice and death would be even more scandalous, thus antitheism should be the natural reaction (Vaas 1999).

Anyway, it is hard to accept for intentional, goal-oriented beings to regard the sometimes sophisticated structures of nature as the result of "blind" self-organized processes of. But exactly this is the scientific approach which cast out any exigencies for design assumptions or teleological explanations. The only opposing trends were some quasi-idealistic interpretations of quantum physics (including the participatory anthropic principle) (Vaas 2001a) and the discussion of the so-called anthropic coincidences or fine-tuning of fundamental constants and some boundary conditions in particle physics and cosmology, sometimes taken as evidence for a cosmic design(er) or teleological force (strong or teleological anthropic principle). Those issues are highly controversial from a scientific and philosophical point of view (Vaas 2004a). But if CAS would be true, basic features of our universe, and even its very existence, could indeed not be explained without reference to intentionality. If cosmological artificial selection was involved, it must be part of such an explanation though it cannot be the full explanation.

Apart from explanatory issues, an attractive psychological feature of CAS might be, at least for some adherents, that it could hold life in high regard. If there is no omnibenevolent god, CAS might point towards a slender substitution after all. So in the face of blank absurdity CAS could be seen as a way out for some deeply frustrated would-be-believers, wanting to restore human grandiosity and an ultimate meaning for everything.

But note that CAS does not necessarily mean that our universe was carefully designed with respect to every law and constant (or specific initial conditions). An engineered (ingenious) blueprint might have been realized, and such ideas are the basic of some "best of all possible worlds" beliefs. But it could also be the case that our universe was just cobbled together, perhaps with many others. Or it might even be an accident, for example in a cosmological or particle collider experiment, i.e. an unintended collateral damage of an otherwise intended action. Although it is hard to imagine, one might even think of many different universes, originated completely naturally, and some of them, including ours, were intentionally picked – like fertilized eggs for uterus implantation in assisted reproductive technology – and activated to develop. (Of course this artificial selection could also have been a purely virtual process within a computer simulation to find out the right world-making recipe, as with iterative numerical calculations if there are no compact analytical solutions – after deriving a successful formula then only the desired universe(s) would have been realized.) Those different possibilities are not mutually exclusive by the way. For example cosmic engineers might create any baby universe – and if it is capable for eternal inflation anything physically possible might ultimately evolve from this. Given the right laws, constants and boundary conditions, even an infinite number of copies (Doppelgänger) of their own would emerge, and any possible variation of them (see below). Thus, as in CNS or eternal inflation scenarios, life and intelligence might be inevitable, although still accidental in some sense. So a kind of radical contingency remains. And of course one can still argue that life is absurd if anything that can happen will happen – and with any possible variance as well and infinitely often so. Indeed exactly that was the point of Friedrich Nietzsche discussing eternal recurrence.

In conclusion, CAS is neither restricted to a careful and complete world design nor does it imply that every law, constant and/or initial condition was intentionally selected. Creating life (let alone human beings!) need not be the goal of this art of world-making either. Perhaps life was just allowed for – or even an accident or mistake. If so, CAS would not prevent (our) life from being incidental. (Though at least we would have someone to blame for all the blunder.)
Even if one accepts CAS, without further knowledge it is impossible to tell anything about the intentions of the creator(s). They might work in mysterious ways their wonders to perform... This is, of course, another problem for CAS. An intentional explanation without explaining the intention might be considered as a shortcoming. But this is not a refutation. And speculations are possible too.
For example it was suggested that the creators – not taken as god(s) but as technologically very advanced, though nevertheless limited cosmic engineers – are simply curious; so they might be trying hard to figure out ways of universe formation (an engineer's proverb states that one only understands something if one is able to make it). Then life might be an accident indeed.
Another possibility is that those cosmic engineers created their own universal soap opera for entertainment (perhaps even with sadistic intentions). Life would not be incidental then, but something like a zoo inhabitant or gladiator. Furthermore, our universe might soon become too boring for its spectators and therefore suddenly be deleted...
Much more serious is the assumption that the cosmic engineers face their own death too, and the forthcoming end of their universe. Thus they might try to escape into a kind of rescue universe or at least transmit something of their knowledge lest they will not be forgotten completely (see below). This brings us back to absurdity. Death might be seen as an ultimately salvation, but it also marks the ultimate absurdity. To fight futility, self-conscious life gets the urge to endure and to intellectually grow endlessly. If this takes infinitely many (infinite?) universes, why not try to make them, if this is possible?

## 3   Is life ultimately doomed?

It is an age-old question, whether the universe is infinite in time and space – or at least part of a larger system which is – and what this means for the ultimate prospects of life. In a branch of modern cosmology, sometimes called *physical eschatology*, this question can be discussed within the framework of (albeit speculative) scientific reasoning (table 2).
The fate of the universe and intelligence depends crucially on the nature of the still mysterious *dark energy* which probably drives the accelerated expansion. Depending on dark energy's – perhaps time-dependent – equation of state, there is now a confusing number of mutually exclusive models. They are popularly called big whimper, big decay, big crunch, big brunch, big splat, big oscillation, big brake, big freeze, big rip, big trip, big hit, big hole, big resurrection etc., and they envisage many different avenues (Vaas 2010b). Most of them are dead ends for life – and this is also true for other cosmological models without dark energy. Thus, the ultimate future of our universe looks deadly dark (Vaas 2006a & 2010b).
But many self-conscious individuals want to fight absurdity and overcome death. If cosmological boundary conditions or creative minds – not necessarily god(s) – beyond it (see e.g. Leslie 2001 & 2008 for a far-reaching proposal), do not support this, mortals must try to take their fate into their own hands, prolonging their life and even searching for a "physics of immortality" (cf. Tipler 1994). Can CAS provide some help here?

*Table 2.* Exploring destiny: Scientific speculations concerning the very far future are bold but not unbound. One must keep some important presuppositions and open problems at the back of one's mind, however. In respect of cosmological artificial selection and artificial cosmogenesis, the possible role of intelligence influencing the fate of the universe is crucial.

| presuppositions for physical eschatology and foundational queries | comments |
|---|---|
| ontological naturalism: nonphysical entities do not exist (or have causal effects); no epistemological idealism or radical illusions about the universe | scientifically unprovable, but strong philosophical arguments in favor |
| ontological status of space and time? | they might be an illusion or not fundamental, but even so predictions and extrapolations are not meaningless and could be rephrased |
| no compact topology of space | otherwise there are different boundary conditions |
| weak determinism (at least); limited effects of chance (acausality) | there might be completely acausal events (like quantum effects), even on macroscopic scales, but no predictions are possible if a weak (moderate, average or statistical) form of determinism doesn't hold |
| relevant laws of nature are known | quite questionable; how will new discoveries change the future and our view of it, are the established scientific methods sufficient at all, and what are the implications of a valid theory of quantum gravity? |
| fundamental laws and constants of nature do not vary | but: possibility of phase transition; perhaps there are no fundamental laws at all (but could effective regularities suffice?); signs of a time-dependent fine-structure constant ($\alpha = e^2/2hce_0 \approx 1/137$) already discovered? |
| problems of infinity | is actual infinity possible in nature? is it realized? how to deal with infinities in theory (calculations, paradoxes) and research (only finite measurements possible)? |
| restricted access? | limited observations in space and time (particle horizon) and finite accuracy of measurements (especially of crucial parameters like $\Omega$ and w) |
| the universe as an open or closed system? | problems for thermodynamics and conservation laws; are there interactions with other universes? |
| limitations of explanations and predictions? | even with (weak) determinism and all relevant laws and boundary conditions known, there might be strong restrictions due to nature's complexity and, perhaps, the incompleteness theorems |
| role of intelligence? | influencing destiny on cosmic scales? |

## 3.1 Dark energy is bad for life

Going along with the externalization of memory and computation through the invention of writing, computers etc., a remarkable, accelerating increase of cultural complexity started on earth (and perhaps at many other places in the universe), a tendency to do ever more work and to require ever less time, space and energy (Fuller 1969, Chaisson 2001, Smart 2008, Vidal 2008). This is an excellent prospect for realizing CAS. However, the accelerated expansion of space leads to an universal limit on the total amount of information that can be stored and processed in the future (Krauss & Starkman 2004): This restricts the technology and computation capabilities for any civilization in principle, because there is access to only a finite volume, even after an infinite time.

For a universe, dominated by a cosmological constant $\Lambda$ (the simplest candidate for dark energy with the energy density $\rho_\Lambda$; for other candidates see e.g. Vaas 2010b), which approaches asymptotically a de Sitter phase where the scale factor a increases exponentially, $a(t) = a_0 e^{Ht}$ with $H = (8\pi\rho_\Lambda/3)^{0.5}$, there is a maximum amount of energy $E_{max}(r)$ that will received by harvesting matter out to a distance r: $E_{max}(r) = \Omega_m c^5/128GH$ where $\Omega_m$ is the matter density, the sum of both baryonic matter (quarks and leptons) and dark matter. $E_{max}(r)$ has a maximum at $Hr/c = 1/2$, the de Sitter horizon is located at $Hr/c = 1$. The accessible energy is only 1/64$^{th}$ of the total energy located within the de Sitter horizon at the present time. With a flat metric (k = 0), a matter density $\rho_m \approx 0.3$, and a Hubble constant $H_0 \approx 70$ kms$^{-1}$Mpc$^{-1}$ one finds $E_{max} \approx 3.5 \cdot 10^{67}$ J. This is comparable to the total rest-mass energy of baryonic matter within today's horizon. Dividing $E_{max}(r)$ by $TK_B\ln 2$, where T is the noise temperature, a minimum energy loss yields a limit on the number of bits that can be processed or information that can be registered. It is smaller than $\pi\Omega_m c^5/64hGH^2\ln 2 = 1.35 \cdot 10^{120}$. (Therefore, by the way, Gordon Moore's law, which assumes that the computer power doubles every 18 to 24 months, cannot continue unabated for more than 600 years for any technological civilization in our observable universe.)

In contrast to a simple eternally expanding universe (big whimper scenario with $\Lambda = 0$), a universe ruled by $\Lambda$ leads to an everlasting expansion with dismal prospects for life. This is due to quantum effects at the cosmic horizon (analogous to Hawking radiation at the horizon of a black hole, but in de Sitter space the horizon surrounds the observer). Because of these the universe cannot cool down to (almost) 0 K. It has a total entropy S and, hence, a final temperature $T_{dS}$ which will be reached within a few hundred billion years (Gibbons & Hawking 1977): $S = A/4 = 3\pi/\Lambda$ and $T_{dS} = 1/2\pi l$ with $A = 4\pi l^2$ and $l = (3/\Lambda)^{0.5}$. Here, A is the area of the de Sitter horizon at late times and l the curvature radius of that closed universe. This de Sitter temperature $T_{dS}$ is approximately $10^{-29}$ K (corresponding to $10^{-33}$ eV). It means the end for any living system because then it cannot radiate away waste heat – and there is no life without an energy gradient (Krauss & Starkman 2000).

Other scenarios look also more or less disappointing. But if our universe and every living being in it would be finally doomed, there could be infinitely many other universes and/or our universe might *recycle* itself due to new inflationary phase transitions out of black holes (Smolin 2010) or out of its high energy vacuum state, where new exponential expanding bubbles should nucleate at a constant rate, growing to new universes elsewhere with new thermalized regions (Lee & Weinberg 1987, Garriga & Vilenkin 1998), and cut their cords, metaphorically speaking. They probably will give rise to new galaxies and civilizations. It is not possible, however, to transcend these boundaries or to send a device with the purpose to recreate a follow-up of the original civilization in the new region or to transmit at least a kind of cosmic message in a bottle. It is not possible (Garriga et al. 2000), because the device or message will almost certainly be intercepted by black holes, which nucleate at a much higher rate than inflating bubbles, namely in the order of $\sim \exp(10^{122})$.

### 3.2 Wormhole escapism and designer universes

If our universe is ultimately determined to die, or if at least the sufficient conditions for any possible information processing system disappear, the only chance for life would be to leave its universe and move to another place. Therein lays the prospect for an everlasting future of civilizations, and this is a strong motivation for CAS. But, as mentioned above, this relocation must happen without quantum tunneling. Because of extremely small tunneling probabilities all mechanisms that involve quantum tunneling are probably doomed to failure. However, there are bold speculations about *traversable wormholes* leading to other universes (Visser 1996, Vaas 2005). This seems to be possible at least in the framework of general relativity.

Perhaps wormholes could be found in nature and modified, or they could be built from scratch. If so, life could switch to another universe, escaping the death of its home.

And if there is no life-friendly universe with the right conditions (physical constants and laws), an advanced civilization might even create a sort of replacement or rescue universe on its own. In fact, some renowned physicists have speculated about such a kind of world-making (Farhi & Guth 1987, Frolov et al. 1989, Farhi et al. 1990, Fischler et al. 1990, Linde 1992, Crane 1994, Harrison 1995, Merali 2006, Ansoldi & Guendelman 2006 & 2008).

At a Grand Unified Theory energy scale of $10^{14}$ GeV a universe might emerge from a classical bubble which starts out with a mass of only about 10 kg. By means of quantum tunneling, the bubble mass could be arbitrarily small, but the formation probability of a new universe would be reduced very much. Of course the main problem is to concentrate enough energy in a tiny volume. It has been suggested to try a coalescence of two regular magnetic monopoles (with below critical magnetic charge), producing a supercritical one which then inflates giving rise to a baby universe, or to take just one monopole and to hurl mass onto it, using a particle accelerator or a cosmic string (Borde et al. 1999, Sakai et al. 2006). The new bubble filled with a false vacuum is an extremely warped patch of spacetime and would create its own space: It undergoes an internal exponential inflation without displacing the space outside of the bubble itself (the negative pressure inside, zero outside, and the positive surface tension prevent the bubble from expanding into its mother universe). On the contrary it disconnects from the exterior region: The wormhole, which acts like an umbilical cord between the mother and child universe, collapses. (From the perspective of the mother universe the disconnected bubble hides inside a microscopic black hole which will not appear to grow in size but evaporates quickly, while from within the bubble the creation event is seen as a white hole-like initial singularity. Mathematically, the bubble can be described as a de Sitter spacetime embedded in a Schwarzschild spacetime, joined by using the Israel junction conditions.) The new universe might be detectable nevertheless because of modifications to the Hawking radiation. It remains unclear, however, whether one could pass a message to the future inhabitants of the created universe (Hsu & Zee 2006) – due to inflation they would live in a tiny corner of a single letter, so to speak. Perhaps it could be encoded within the value of a fundamental constant. It remains also unclear, whether one could even travel into the descendent universe via new wormholes. If such an interchange of universes is possible, life might continue endlessly.

While such bold speculations might sound awkward or technocratic or as the ultimate megalomania, they at least offer an interesting change of perspective (from observation to experiment), which questions the passive point of view when dealing with cosmological problems and the limits of observations due to the restrictions imposed by the spacetime structure on the causal relations among objects. This is another advantage of CAS.

Like dark energy, however, wormholes violate some fundamental energy conditions. And a violation of the weak energy condition (WEC) is also necessary to create new inflating regions without quantum tunneling and to go there or send messages into it, for instance a blueprint of the engineering civilization. The required magnitude of the negative energy density is in the order of $-\rho = H_{inf}^{-2}$, where $H_{inf}$ is the inflationary expansion rate. Because WEC violation is in conflict with quantum inequalities (Ford & Roman 1997, Borde at al. 2002, Ford et al. 2002), it should be investigated how seriously this constraint is to be taken, since it is unclear to what extent these inequalities apply to interacting fields.

### 3.3 Eternal recurrence and the ultimate Copernican principle

Accepting the cosmological constraints, it seems to be very unlikely that a civilization can survive forever within its universe. In the end it will either be crushed down, ripped apart or it will run out of energy. But the latter is a statistical argument, since it is based on the second law of thermodynamics. Entropy can decrease due to thermal *fluctuations*, and it is in principle possible that such fluctuations can sustain a civilization for an arbitrarily long time – at least if there are infinitely many in an infinite, future-eternal space. The probability that a civilization will survive for some time is a sharply decreasing function of that time. However, for any finite time the probability is finite, and thus many civilizations will live longer than any given time, if the universe is large enough to allow those improbable events to occur. Of course, there is almost no chance that our own successors are going to be among the lucky civilizations whose life will be prolonged by thermal fluctuations in such a lottery universe.

Furthermore, in an infinite future time might not be a problem. Eventually, anything could spontaneously pop into existence due to quantum fluctuations if spacetime is eternal. They would mostly result in meaningless garbage, but a vanishingly small proportion shall be people, planets and parades of galaxies. This paper will reappear again too. And such a kind of quantum resurrection might even spark a new big bang. According to Sean Carroll and Jennifer Chen (2004) one must be patient, however, and wait some $10^{10^{56}}$ years (if a stable de Sitter vacuum is the "natural ground state"). Our whole universe might be such an island in a Λ-sea (Dutta & Vachaspati 2010).

But on the other hand there are deep problems with fluctuation scenarios, because it is much more likely that, for instance, a conscious system pops out of the vacuum for a few seconds, and everything "around us" (including this paper) is its mere solipsistic illusion (Albrecht & Sorbo 2004, De Simone et al. 2008, Vaas 2009d). To give some thermodynamical numbers of entropy fluctuations in a de Sitter background: The probability of our observable universe is just $1:10^{10^{123}}$ – extremely tiny in contrast to a spontaneous ex nihilo origination of a freak observer: Perhaps $1:10^{10^{21}}$ for the smallest possible conscious computer and between $1:10^{10^{51}}$ and $1:10^{10^{70}}$ for a "Boltzmann brain". (The probability for a solar system like ours coming out of the vacuum is $1:10^{10^{85}}$.) Thus, there is a controversial discussion going on about wrong assumptions underlying those kinds of estimates – not because scientists believe that such a solipsistic illusion is true but because these probabilities indicate possible errors in cosmological reasoning and deep difficulties of multiverse models, especially the measure problem in inflationary cosmology (Bousso et al. 2006, Aguirre et al. 2007, Vilenkin 2007, Linde 2007, Li & Wang 2007, Garriga & Vilenkin 2009, Linde et al. 2009).

Disregarding fluctuation issues, something strange is inevitable nevertheless, if two conditions are true: firstly, if our universe is infinite, or if there are infinitely many other universes with the same laws and constants; and, secondly, if quantum theory holds and there is, therefore, a finite number of possible states (that is due to Heisenberg's uncertainty relation there is no continuum of states and perhaps not even of space and time). If those two assumptions are valid, then according to Alexander Vilenkin, Jaume Garriga and others every combination of discrete finite physical states are realized arbitrarily or infinitely often (Ellis & Brundrit 1979, Garriga & Vilenkin 2001, Vaas 2001b, Knobe et al. 2006, Vilenkin 2006). (Imagine a lattice built randomly out of zeros and ones: Every finite combination of zeros and ones, that is every "local" pattern occurs infinitely often.) Thus, there is a kind of spatial eternal recurrence.

This also implies that we would have perfect copies: *Doppelgänger* which are identical to us as far as quantum physics allows, and also Doppelgänger biographies, Doppelgänger earths, solar systems, milky ways and even Hubble volumes. Their distances are vast, but not infinite, and they could even be estimated, as Max Tegmark did: Our personal neighbouring Doppelgänger should be $10^{10^{29}}$ m apart, and another Doppelgänger Hubble volume, that is a region of space exactly like our observable universe, $10^{10^{115}}$ m (Tegmark 2004). This spatial eternal recurrence could extend in time, which seems to be true either in a future-eternal inflationary or cyclic scenario with a flat universe. Thus, even if the history of our universe (and/or every universe) might lead to a global death, everything and every life-form might reappear over and over again, infinitely often both in space and time. Then, it is true that there is no personal eternal life, because every organism is doomed, but life as such could not be driven out of existence completely everywhere and everywhen. It would be truly eternal.

On the other hand, eternal recurrence seems to be absurd. And it is not only exact duplication – it is also every possible alternative, because all variations are equally real (note that this has nothing to do with the many worlds interpretation of quantum mechanics, although it is compatible with it). As Alexander Vilenkin has said, some people "will be pleased to know that there are infinitely many […] regions where Al Gore is President and – yes – Elvis is still alive". Thus, physical potentiality and actuality would ultimately be the same. If so, the search for options and the struggle for life doesn't matter globally. Everything that might happen will happen at one place or another – in fact, it will happen infinitely often. This can be seen as the *ultimate Copernican principle.* It could be disappointing or encouraging, depending on personal taste. However, it seems very strange and for many people even insulting, that we are not unique, and that everything we try might succeed here but not elsewhere and vice versa – infinitely often.

If the ultimate Copernican principle or any of the Doppelgänger scenarios is true, the urge for artificial cosmogenesis and a need for CAS are superfluous. Life would not *continue* forever, but can *exist* endlessly nevertheless. Everything will repeat itself eternally – including CAS ideas and proponents.

## 4      The case for CAS? – Conclusions and outlook

The hypothesis of cosmological artificial selections does not only address (1) the origin and apparent fine-tuning of our universe but also (2) the possible value and meaning of life and (3) its ultimate future. However, all these complex issues provide eminent problems for CAS. One might argue that although CAS is based on three weak points, putting them together they make the case for CAS stronger, i.e. strengthen its stability under load like a tripod. Indeed, cosmic fine-tuning, meaning and survival are fancily linked together in the CAS scenario and form a coherent picture. But this does not make the CAS proposal true, of course. And, indeed, the three points are qualitatively distinct: Fine-tuning is about explanation, meaning about evaluation, and survival about action and construction. Therefore it is questionable whether one can really strengthen the others, although explanation might be a necessary condition for construction (or vice versa?) and (the search for) meaning a crucial motivation for explanation and action. Nevertheless, all three points and, thus, CAS remain an open issue at the moment.

CAS is (or a least starts out as) a metaphysical speculation. And there is nothing wrong with metaphysical speculations if they are not confused with or advertised as scientific results. What's more, (some) metaphysical speculations have a heuristic value and might even boost the formation of scientific hypotheses. And philosophy is, among other things, thinking in advance. Both the challenge of escaping cosmic doomsday and searching for penultimate explanations – really ultimate explanations are excluded (Vaas 2006b) – surely need unconventional input and encouragement. But CAS is or can be seen also as a scientific speculation. Like multiverse scenarios in general, it fulfills many criteria of science (Vaas 2008c) and could even be testable – or realizable – in the future. CAS might be judged as unlikely or far-fetched, but it is worth exploring. It extends the realm of both cosmological problems and possible solutions and, thus, challenges other approaches – constructive competition is always good for science and philosophy, and criticism is a gift for further developments.

Summing up, it seems quite unlikely that the hypothesis of cosmological artificial selection is correct (at least as the cause of our universe): First there are other more likely and simpler explanations for the fine-tuning of our universe (or for getting rid of the anthropic coincidences altogether); second psychological urges for overcoming human contingency are no argument for the truth of scientific hypotheses, and CAS is far from being an analgesic against absurdity; and third it seems unlikely that an advanced civilization within our universe can intentionally start the creation of new universes either by simulating them (because of the finite computational resources both in size and in time) or by physically producing them (because this might either be too difficult or it happens naturally much earlier and more often anyway).

However, if CAS is possible in principle, our successors or any other much further advanced civilization within our universe might be the very first to fully realize it nevertheless. If this occurs as a simulation or emulation, its contents – as complex as they might be, perhaps including even self-conscious beings – ultimately would be doomed if the simulating hardware breaks down. And within our universe, this seems to be inevitable. Thus, such simulated universes cannot endure for ever. (If we ourselves would live within a computer simulation, or rather be one, the show might stop very soon... without any prospect for a cosmic reset.) If, on the other hand, somebody within our universe can artificially create offspring universes and even transmit the recipe for doing that – either as a message or as a physical necessity for instance by starting Doppelgänger universes which inevitably will repeat history – then a potentially infinite chain of successor universes might begin. Eternal life, then, becomes a reality, even if it is not necessarily an eternal *continuing* life.

Assuming that such a giant chain of being is actually possible, however, it seems nevertheless quite unlikely that our universe is the very first one to accomplish this. Furthermore, this would be a violation of the Copernican principle because our location in spacetime, in this case the multiverse, would be very special. Therefore one should conclude that, given the CAS framework would be correct indeed, our universe is a result of cosmological artificial selection (or simulation) too – one link within the probably future-eternal chain. If so, the spark of life may endure endlessly indeed. And even if we or our successors would not be able to pass it on, being then a tiny dead end within a flourishing realm of evolution, we will at least have envisioned it.

**Acknowledgements**

I am grateful to Anthony Aguirre, Juan García-Bellido, John Leslie, Andrei Linde, Lee Smolin, Paul Steinhardt, and Alex Vilenkin for discussion over the years as well as Angela Lahee, Nela Varwig and especially André Spiegel for their kind support. Thanks also to Clément Vidal for motivation and the invitation to comment. – Scientific speculation and philosophy of science and nature are often dangerous fields, but useful and thrilling nevertheless for getting ideas, criticism and motivation to struggle against the boundaries of experience, empirical research, established theories, and imagination. As Carl Sandburg once wrote: "Nothing happens unless first a dream."

**References**

Aguirre, A. (2010): Eternal Inflation: Past and Future. In: Vaas, R. (ed.) (2010): Beyond the Big Bang. Springer, Heidelberg.
Aguirre, A., Gratton, S. Johnson, M.C. (2007): Hurdles for recent measures in eternal inflation. Phys. Rev. D 75, 123501; arXiv:hep-th/0611221
Albrecht, A., Sorbo, L. (2004): Can the universe afford inflation? Phys. Rev. D 70, 063528; arXiv:hep-th/0405270
Ansoldi, S., Guendelman, E.I. (2006): Child Universes in the Laboratory; arXiv:gr-qc/0611034
Ansoldi, S., Guendelman, E.I. (2008): Universes out of almost empty space. Prog. Theor. Phys. 120, 985–993; arXiv:0706.1233
Barrabès, C., Frolov, V.P. (1996): How many new worlds are inside a black hole? Phys. Rev. D 53, 3215–3223; arXiv:hep-th/9511136
Borde, A., Ford, L.H., Roman, T.A. (2002): Constraints on Spatial distributions of Negative Energy. Phys. Rev. D 65, 084002; arXiv:gr-qc/0109061
Borde, A., Trodden, M., Vachaspati, T. (1999): Creation and Structure of Baby Universes in Monopole Collisions. Phys. Rev. D 59, 043513; arXiv:gr-qc/9808069
Bostrom, N. (2003): Are We Living in a Computer Simulation? Phil. Quart. 53, 243–255.
Bousso, R., Freivogel, B., Yang, I.S. (2006): Eternal Inflation: The Inside Story. Phys. Rev. D 74, 103516; arXiv:hep-th/0606114
Byrne, P. (1989): Natural religion and the nature of religion. Routledge, London.
Carr, B. (2007): The Anthropic Principle Revisited. In: Carr, B. (ed.) (2007): The Universe or Multiverse? Cambridge University Press, Cambridge, pp. 77–89.
Carroll, S.M., Chen, J. (2004): Spontaneous Inflation and the Origin of the Arrow of Time; arXiv:hep-th/0410270
Chaisson, E.J. (2001): Cosmic Evolution: The Rise of Complexity in Nature. Harvard University Press, Cambridge.
Crane, L. (1994): Possible Implications of the Quantum Theory of Gravity; arXiv:hep-th/9402104
Davidson, D. (2001): Essays on Actions and Events. Oxford University Press, Oxford.
Davies, P. (2007): Universes galore: where will it all end? In: Carr, B. (ed.) (2007): The Universe or Multiverse? Cambridge University Press, Cambridge, pp. 487–505.
De Simone, A. et al. (2008): Boltzmann brains and the scale-factor cutoff measure of the multiverse; arXiv:0808.3778
Deism. Wikipedia (accessed 29 October 2009), http://en.wikipedia.org/wiki/Deism
Dennett, C. (1995): Darwin's Dangerous Idea. Simon & Schuster, New York.
Douglas, M. (2010): The String Landscape: Exploring the Multiverse. In: Vaas, R. (ed.) (2010): Beyond the Big Bang. Springer, Heidelberg.
Duff, M.J., Okun, L.B., Veneziano, G. (2002): Trialogue on the number of fundamental constants. JHEP 0203, 023; arXiv:physics/0110060
Dutta, S., Vachaspati, T. (2010): Island Cosmology: The Universe from a Quantum Fluctuation. In: Vaas, R. (ed.) (2010): Beyond the Big Bang. Springer, Heidelberg.
Ellis, G.F.R., Brundrit, G.B. (1979): Life in the infinite universe. Quart. J. Royal. Astr. Soc. 20, 37–41.
Ellis, G. (1997): A Darwinian universe? Nature 387, 671–672.
Farhi, E., Guth, A.H. (1987): An obstacle to creating a universe in the laboratory. Phys. Lett. B 183, 149–155.
Farhi, E., Guth, A.H., Guven, J. (1990): Is it possible to create a universe in the laboratory by quantum tunnelling? Nucl. Phys B 339, 417–490.
Fischler, W., Morgan, D., Polchinski, J. (1990): Quantum nucleation of false-vacuum bubbles. Phys. Rev. D 41, 2638–2641.
Ford, L.H., Roman, T.A. (1997): Restrictions on Negative Energy Density in Flat Spacetime. Phys. Rev. D 55, 2082–2089; arXiv:gr-qc/9607003
Ford, L.H., Helfer, A.D., Roman, T.A. (2002): Spatially Averaged Quantum Inequalities Do Not Exist in Four-Dimensional Spacetime. Phys. Rev. D 66, 124012; arXiv:gr-qc/0208045
Frolov, V.P., Markov, M.A., Mukhanov, M.A. (1989): Through a black hole into a new universe? Phys. Lett. B 216, 272–276.
Fuller, R.B. (1969): Utopia Or Oblivion: The Prospects for Humanity. Overlook Press, New York.

García-Bellido, J. (1995): Quantum Diffusion of Planck Mass and the Evolution of the Universe. In: Occhionero, F. (ed.) (1995): Birth of the Universe and Fundamental Physics. Lecture Notes in Physics 455. Springer, Berlin, pp. 115–120; arXiv:astro-ph/9407087

Garriga, J., Vilenkin, A. (1998): Recycling universe. Phys. Rev. D 57, 2230–2244; arXiv:astro-ph/9707292.

Garriga, J., Vilenkin, A. (2001): Many worlds in one. Phys. Rev. D 64, 043511; arXiv:gr-qc/0102010

Garriga, J., A. Vilenkin, A. (2009): Holographic Multiverse. JCAP 0901, 021; arXiv:0809.4257

Garriga, J. et al. (2000): Eternal inflation, black holes, and the future of civilizations. Int. J. Theor. Phys. 39, 1887–1900; arXiv:astro-ph/9909143

Gasperini, M., Veneziano, G. (2010): The Pre-Big Bang Scenario: String Theory and a Longer History of Time. In: Vaas, R. (ed.) (2010): Beyond the Big Bang. Springer, Heidelberg.

Gay, P. (ed.) (1968): Deism. Van Nostrand, Princeton.

Gibbons, G., Hawking, S.W. (1977): Cosmological Event Horizons, Thermodynamics, and Particle Creation, Phys. Rev. D 15, 2738–2751.

Hartle, J., Hawking, S.W. (1983): The wave function of the universe. Phys. Rev. D 28, 2960–2975.

Harrison, E.R. (1995): The natural selection of universes containing intelligent life. Quart. J. Royal Astr. Soc. 36, 193–203.

Hempel, C.G. (1965): Aspects of Scientific Explanation and Other Essays in the Philosophy of Science. Free Press, New York.

Hogan, C.J. (2000): Why the Universe is Just So. Rev. Mod. Phys. 72, 1149–1161; arXiv:astro-ph/9909295

Hsu, S., Zee, A. (2006): Message in the Sky. Mod. Phys. Lett. A 21, 1495–1500; arXiv:physics/0510102

Johnson, B. (2009): Deism. Truth Seeker, Escondido.

Kane, G.L., Perry, M J., Zytkow, A.N. (2002): The Beginning of the End of the Anthropic Principle. New Astron. 7, 45–53; arXiv:astro-ph/0001197

Kanitscheider, B. (2009): Darwins Theorie als Prototyp und Vorläufer einer Theorie der Selbstorganistion. Universitas 64, 56–66.

Knobe, J., Olum, K.D., Vilenkin, A. (2006): Philosophical Implications of Inflationary Cosmology. Brit. J. Phil. Sci. 57, 47–67; arXiv:physics/0302071

Krauss, L.M., Starkman, G.D. (2000): Life, The Universe, and Nothing. Astrophys. J. 531, 22–30; arXiv:astro-ph/9902189

Krauss, L.M., Starkman, G.D. (2004): Universal Limits on Computation; arXiv:astro-ph/0404510

Lee, K.M., Weinberg, E.J. (1987): Decay Of The True Vacuum In Curved Space-Time. Phys. Rev. D 36, 1088–1094.

Leslie, J. (1989): Universes. Routledge, London 1996.

Leslie, J. (2001): Infinite Minds. Clarendon Press, Oxford.

Leslie, J. (2008). Infinitely Long Afterlives and the Doomsday Argument. Philosophy 83, 519–524.

Li, M., Wang, Y. (2007): A Stochastic Measure for Eternal Inflation. JCAP 0708, 007; arXiv:0706.1691

Linde, A.D. (1987): Particle physics and inflationary cosmology. Phys. Today 40 (9), 61–68.

Linde, A. (1992): Hard Art of the Universe Creation. Nucl. Phys. B 372, 421–442; arXiv:hep-th/9110037

Linde, A. (2005): Particle Physics and Inflationary Cosmology. Contemp.Concepts Phys. 5 1–362; arXiv:hep-th/0503203

Linde, A. (2007): Towards a gauge invariant volume-weighted probability measure for eternal inflation. JCAP 0706, 017; arXiv:0705.1160

Linde, A. (2008): Inflationary Cosmology. Lect. Notes Phys.738, 1–54; arXiv:0705.0164

Linde, A., Vanchurin, V. (2009): How many universes are in the multiverse? arXiv:0910.1589

Linde, A., Vanchurin, V., Winitzki, S. (2009): Stationary Measure in the Multiverse. JCAP 0901, 031; arXiv:0906.4954

Lipton, P. (2004): Inference to the Best Explanation. Routledge, London, 2. ed.

Mayes, G. R. 2005: Theories of Explanation. The Internet Encyclopedia of Philosophy. http://www.utm.edu/research/iep/e/explanat.htm

Maynard Smith, J., Szathmáry, E. (1996): On the likelihood of habitable worlds. Nature 384, 107.

Merali, Z. (2006): Create your own universe. New Scientist 2559, 32–35.

Mersini-Houghton, L. (2010): Selection of Initial Conditions: The Origin of Our Universe from the Multiverse. In: Vaas, R. (ed.) (2010): Beyond the Big Bang. Springer, Heidelberg.

Monod, J. (1970): Chance and Necessity. Knopf, New York 1971.

Pitt, J.C. (ed.) (1988): Theories of Explanation. Oxford University Press, New York.

Rothman, T., Ellis, G.F.R. (1993): Smolin's natural selection hypothesis. Quart. J. Royal Astr. Soc. 34, 201–212.

Sakai, N. et al. (2006): The universe out of a monopole in the laboratory? Phys. Rev. D 74, 024026; arXiv:gr-qc/0602084

Salmon, W.C. (1998): Causality and explanation. Oxford University Press, New York.

Smart, J. (2008): Evo Devo Universe? A Framework for Speculations on Cosmic Culture. In: Dick, S.J., Lupisella, M. (eds.) (in print): Cosmos and Culture. http://accelerating.org/downloads/SmartEvoDevoUniv2008.pdf

Smolin, L. (1992): Did the universe evolve? Class. Quant. Grav. 9, 173–191.

Smolin, L. (1997): The Life of the Cosmos. Oxford University Press, Oxford.

Smolin, L. (2010): Cosmological Natural Selection: Status and Implications. In: Vaas, R. (ed.) (2010): Beyond the Big Bang. Springer, Heidelberg.

Tegmark, M. et al. (2006): Dimensionless constants, cosmology, and other dark matters. Phys. Rev. D 73, 23505; arXiv:astro-ph/0511774

Tegmark, M. (2004): Parallel Universes. In: Barrow, J., Davies, P.C.W., Harper jr., C.L. (eds.) (2004): Science and Ultimate Reality. Cambridge University Press, Cambridge, pp. 459–491; arXiv:astro-ph/0302131

Tegmark, M. (2010): The Mathematical Universe: Eternal Laws and the Illusion of Time. In: Vaas, R. (ed.) (2010a): Beyond the Big Bang. Springer, Heidelberg.

Tipler, F.J. (1994): The Physics of Immortality. Anchor Books, New York.

Vaas, R. (1993): Die Welt als Würfelspiel. In: Evangelische Akademie Baden (ed.) (1993): "Gott würfelt (nicht)!" Karlsruhe, pp. 108–162.

Vaas, R. (1995a): Reduktionismus und Emergenz. In: Die mechanische und die organische Natur. Beiträge zum Naturverständnis. Konzepteheft 45 des SFB 230. Stuttgart, Tübingen 1995, pp. 102–161.

Vaas, R. (1995b): Masse, Macht und der Verlust der Einheit. In: Krüger, M. (ed.) (1995): Einladung zur Verwandlung. Hanser. München, pp. 219–260.

Vaas, R. (1998): Is there a Darwinian Evolution of the Cosmos? – Some Comments on Lee Smolin's Theory of the Origin of Universes by Means of Natural Selection. Proceedings of the MicroCosmos – MacroCosmos Conference, Aachen; arXiv:gr-qc/0205119

Vaas, R. (1999): Der Riß durch die Schöpfung. der blaue reiter. Journal für Philosophie 10, 39–43.

Vaas, R. (2001a): Why Quantum Correlates Of Consciousness Are Fine, But Not Enough. Informação e Cognição 3 (1), 64–107. http://www.portalppgci.marilia.unesp.br/reic/viewarticle.php?id=16

Vaas, R. (2001b): Ewiges Leben im Universum? bild der wissenschaft 9, 62–67.

Vaas, R. (2003): Problems of Cosmological Darwinian Selection and the Origin of Habitable Universes. In: Shaver, P.A., DiLella, L., Giménez, A. (eds.): Astronomy, Cosmology and Fundamental Physics. Springer, Berlin, pp. 485–486.

Vaas, R. (2004a): Ein Universum nach Maß? Kritische Überlegungen zum Anthropischen Prinzip in der Kosmologie, Naturphilosophie und Theologie. In: Hübner, J., Stamatescu, I.-O., Weber, D. (eds.) (2004): Theologie und Kosmologie. Mohr Siebeck, Tübingen, pp. 375–498.

Vaas, R. (2004b): Time before Time. Classifications of universes in contemporary cosmology, and how to avoid the antinomy of the beginning and eternity of the world. arXiv:physics/0408111

Vaas, R. (2005): Tunnel durch Raum und Zeit. Kosmos, Stuttgart.

Vaas, R. (2006a): Dark Energy and Life's Ultimate Future. In: Burdyuzha, V. (ed.) (2006): The Future of Life and the Future of our Civilization. Springer, Dordrecht, pp. 231–247. arXiv:physics/0703183

Vaas, R. (2006b): Das Münchhausen-Trilemma in der Erkenntnistheorie, Kosmologie und Metaphysik. In: Hilgendorf, E. (ed.) (2006): Wissenschaft, Religion und Recht. Logos, Berlin, pp. 441–474.

Vaas, R. (2008a): Hawkings neues Universum. Wie es zum Urknall kam. Kosmos, Stuttgart.

Vaas, R. (2008b): Aufrechtstehen im Nichts. Universitas 63, 1118–1137 & 1244–1259.

Vaas, R. (2008c): Phantastische Physik: Sind Wurmlöcher und Paralleluniversen ein Gegenstand der Wissenschaft? In: Mamczak, S., Jeschke, W. (eds.): Das Science Fiction Jahr 2008. Heyne, München, pp. 661–743.

Vaas, R. (2009a): Die Evolution der Evolution. Universitas 64, 4–29.

Vaas, R. (2009b): Gods, Gains, and Genes. On the Natural Origin of Religiosity by Means of Bio-cultural Selection. In: Voland, E., Schiefenhövel, W. (eds.) (2009): The Biological Evolution of Religious Mind and Behavior. Springer, Heidelberg, pp. 25–49.

Vaas, R. (2009c): Götter, Gene und Gehirne – Biologische Grundlagen der Religiosität. Die Kunde (in press).

Vaas, R. (2009d): Das wahnsinnige Universum. bild der wissenschaft 3, 58–61.

Vaas, R. (ed.) (2010a): Beyond the Big Bang. Springer, Heidelberg

Vaas, R. (2010b): Eternal Existence. In: Vaas, R. (ed.) (2010): Beyond the Big Bang. Springer, Heidelberg.

Vidal, C. (2008): The Future of Scientific Simulations: from Artificial Life to Artificial Cosmogenesis. In: Tandy, C. (2008): Death And Anti-Death. Ria University Press, Palo Alto, pp. 285–318; arXiv:0803.1087

Vidal, C. (2010): Computational and Biological Analogies for Understanding Fine-Tuned Parameters in Physics. Foundations of Science. http://evodevouniverse.com/EDU2008Papers/VidalBiologicalComputationalAnalogiesFineTuningEDU2008.pdf

Vilenkin, A. (1995): Predictions from Quantum Cosmology. Phys. Rev. Lett. 74, 846–849; arXiv:gr-qc/9406010

Vilenkin, A. (2006): Many Worlds in One. Hill and Wang, New York.

Vilenkin, A. (2007): A measure of the multiverse. J. Phys. A 40, 6777–6785; arXiv:hep-th/0609193

Visser, M. (1996): Lorentzian Wormholes. American Institute of Physics Press, Woodbury.

Waring, E.G. (ed.) (1967): Deism and Natural Religion. Frederick Ungar, New York.

Weinstein, S., Fine, A. (1998): Book Review of Lee Smolin's The Life of the Cosmos. J. Phil. XCV, 264–268.

Wheeler, J.A. (1980): Law without law. In: Medawar, P., Shelley, J. (eds.) (1980): Structure in Science. Elsevier, New York, pp. 132–154.

Wheeler, J.A. (1983): On recognizing law without law. Am. J. Phys. 51, 398–404.

Woodward, J. (2003/2009): Scientific Explanation. Stanford Encyclopedia of Philosophy. http://plato.stanford.edu/entries/scientific-explanation/

Wittgenstein, L. (1922): Tractatus Logico-Philosophicus. Kegan Paul, Trench, Trubner & Co., London.

*October 2009*